# Active coherent beam combining and beam steering using a spatial mode multiplexer

ROMAIN DEMUR*, LUC LEVIANDIER, ELSA TURPIN, JEROME BOURDERIONNET, ERIC LALLIER

*Thales Research and Technology, 1 avenue Augustin Fresnel, 91767 Palaiseau Cedex, France*
**romain.demur@thalesgroup.com*

**Abstract:** Coherent beam combination is one promising way to overcome the power limit of one single laser. In this paper, we use a Multi-Plane Light Converter to combine 12 fibers at 1.03 µm with a phase locking setup. The overall loss measurement gives a combination efficiency in the fundamental Hermite-Gaussian mode as high as 70%. This setup can generate the fundamental and higher-order Hermite-Gaussian modes and has beam steering capabilities.

## 1. Introduction

A wide field of applications ranging from material processing, high-energy physics to laser weapon and free-space optical communication require lasers delivering high brightness optical power. The energy density damage limit in optical materials and the non-linear effects in fibers restrict the maximum achievable optical power, and single lasers with tens of kW average power are still a technological challenge. One promising way to overcome these limits is to combine coherently a large number of fiber amplifiers where one master laser oscillator seeds $N$ optical amplifiers after being split into $N$ independent channels. The outputs of these amplifiers are eventually coherently combined into one single beam using a phase correction method to adjust the phase of each single beam resulting in high power laser beam with good beam quality [1].

Coherent beam combining has been the topic of tremendous works [2] and can be achieved either in tiled-aperture [3] or filled-aperture [4] configuration. The tiled-aperture approach proved his scaling potential but has the drawback of a limited beam combination efficiency due to the far-field diffraction pattern of the output optical head leading to grating lobes and exhibits a theoretical limit of 67% efficiency [3] for Gaussian beams and hexagonal lens array. Despite his combination efficiency limitation, this method exhibits steering capacities – an important feature for precise target or satellite tracking.

On the other hand, with the filled-aperture configuration it is possible to achieve a diffraction limited output beam with a combination efficiency close to 100% but the use of transmissive optics – such as diffractive optical elements – ultimately limits the scalability of the system due to residual absorption of the optics, reducing the maximum achievable power. Finally, beam steering is not achievable using this method.

In this article, we present a coherent beam combination experiment using a non-mode-selective multiplexer based on a multi-plane light conversion (MPLC) device. A MPLC is a spatial multiplexer performing a unitary matrix transformation between two spatial mode basis using a combination of phase plates and free space propagation [5, 6]. For example, a MPLC multiplexer transforms a basis of $N$ spatially separated beams into $N$ spatial modes in a dense basis like a Hermite-Gaussian (HG) or Laguerre-Gaussian basis. This concept has already shown great interest in optical fiber communication with a demonstration using up to 55 spatial modes to obtain a total data rate of 1.53 peta-bit/s [7]. Furthermore, the scalability of the method with 1035 modes [8] and a full spatiotemporal and polarisation control of the light [9] have been experimentally demonstrated showing the device potential for manifold applications. All these MPLCs are mode selective that is, each separate input mode corresponds to a specific output mode.

Recently, non-mode selective MPLCs have been introduced [10, 11] where each specified output mode is a superposition of all the input beams. In this configuration, coherent beam combination is achievable. This specific device transforms a tiled-aperture head into a diffraction limited Gaussian beam using reflective phase plates and mirrors and theoretically leverage the drawbacks of the aforementioned methods. The theoretical coherent combination efficiency is 100%, using a scalable method and reflective-only optics and demonstrates steering capacities. In [10], authors numerically study the coherent beam combining performances of several MPLC design using either Hadamard or Fourier transfer matrix with a reduced number of phase masks. They demonstrate promising results with 256 HG output modes. The first experimental demonstration performed in [11] used 7 modes MPLC at 1.5 µm wavelength and using the LOCSET technique for the phase locking. These authors obtain 72% global combination efficiency.

In our experiment, we use a MPLC from Cailabs [6] with 12 input beams and a phase locking setup to generate the output beam into the fundamental Gaussian mode. We analyse the different loss sources and we then use the MPLC matrix properties to obtain the higher orders of the output HG modes. Finally, we evaluate the beam steering performances of the device.

## 2. Experimental setup

We used a $N = 12$ input fibers MPLC component designed to combine these fibers into a free-space fundamental Gaussian ($HG_{00}$) mode. Fig. 1 illustrates the principle scheme of the device, in (a) the output is the coherent superposition of the input fibers with random phase relationship and (b) is the beam obtained with phase-locking on a uniform phase distribution.

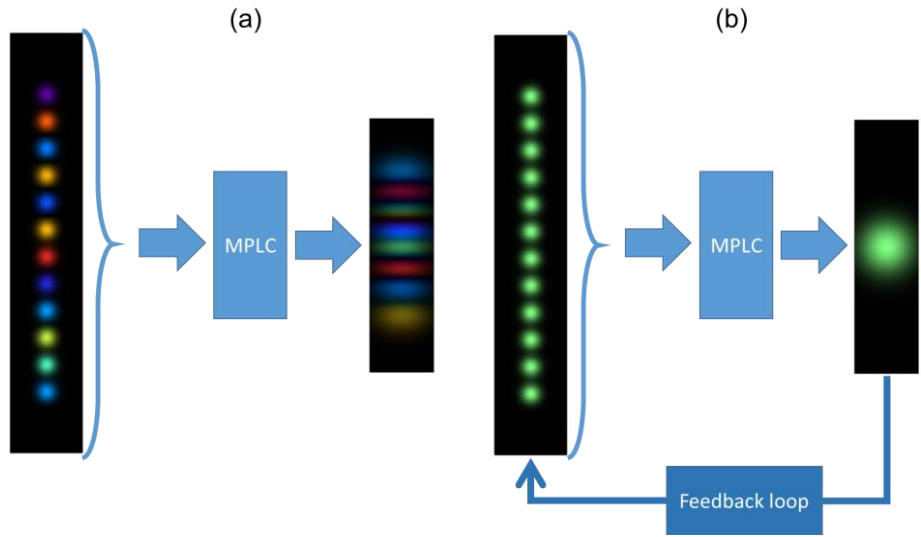

Fig. 1. Principle of a MPLC-based coherent beam combination with 12 input beams. (a) Output example with no phase locking. (b) Output beam with phase locking on the $HG_{00}$ mode. Beam color represents the phase and saturation the intensity.

This transformation is theoretically lossless. However, some losses arise from the reflections on the phase plates, the fiber output coupling, the finite number of phase plates and the device alignment imperfections, resulting in a non-perfect basis transformation.

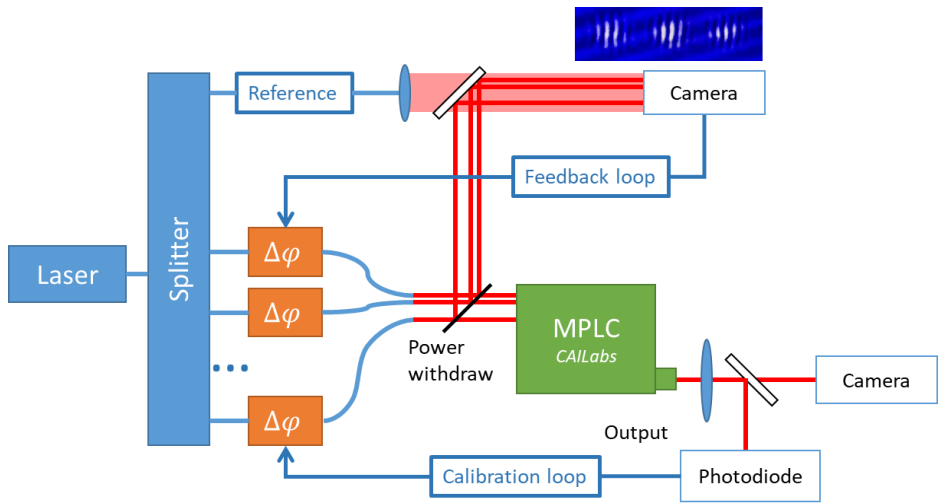

Fig. 2. Scheme of the phase-loop experiment using a MPLC.

Fig. 2 shows our experimental setup. A 1030 nm polarization maintaining (PM) distributed feedback laser (DFB) is split and injected into 12 PM fiber piezo-electric phase shifters. The interferometric phase measurement method described in [12] performs the coherent beam combination. The twelve phase shifted fibers feed the MPLC whereas a 13$^{th}$ output of the splitter is used as reference arm to perform the interferometric images. Inside the MPLC, just after the 1D micro-lensed fiber array, a plate reflects about 1% of the beam power, is near-field imaged and superimposed with the reference arm on a high-speed camera for real-time phase measurement. At the MPLC output, a lens images the beam on a camera and a glass plate samples a part of the power to send it on a photodiode. Using this signal, a Nelder-Mead algorithm focuses the beam into the fundamental $HG_{00}$ mode through feedback on the phase shifters. This optimisation calibrates the phase relationship to apply to the modulators.

## 3. Combination efficiency

Fig. 3 displays the experimental output beam before (a) and after (b) the Nelder-Mead optimization. We notice a beam combination qualitatively close to a Gaussian beam. However, the beam is slightly flattened and the result is not exactly the fundamental mode. Fig. 3 (c) represents the power in the bucket of the combined beam as a function of the encircled radius in blue line compared to a perfect Gaussian beam in red line. For a radius of one waist, the power in the bucket is 89% of the one expected in theory.

This good figure of combination efficiency does not take into account all the loss sources. One source of loss is the input/output power transmission of the MPLC, measured at 93% transmission, after subtraction of fiber connector loss estimated at -0.5 dB.

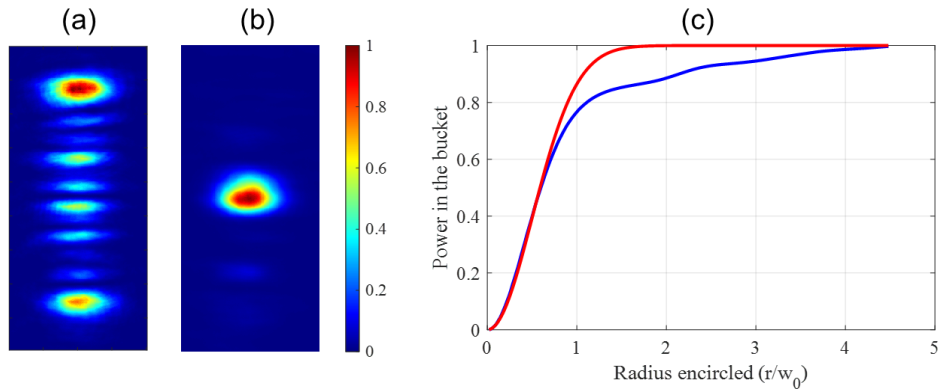

Fig. 3. (a) Experimental example of an output beam without and (b) with phase-locking. (c) Power in the bucket as a function of circle radius for the beam with phase locking (blue) and a comparison with a perfect Gaussian beam (red).

The last source of loss is the light scattering on phase plates coming out of the MPLC outside of the spatial support of the 12 first HG modes. Fig. 4 exhibits this behaviour where the output intensity is plotted in log scale. There is a speckle-like loss around the spatial limit of the $HG_{0,11}$ mode support represented by a black rectangle due to imperfections in phase plate fabrication and components alignment in the MPLC. We evaluate the loss from this source to 16%. This value is higher than the one reported in [11]. The main difference between the two devices is the operational wavelength: 1.55 µm compared to 1.03 µm in our case. As the Rayleigh scattering on phase plates and mirrors scales as $1/\lambda^4$, with $\lambda$ the optical wavelength, we suffer from a scattering 5 times greater than in [11]. The MPLC device loss is highly sensitive to misalignment error and a slightly lower alignment quality may also contribute to the higher loss observed.

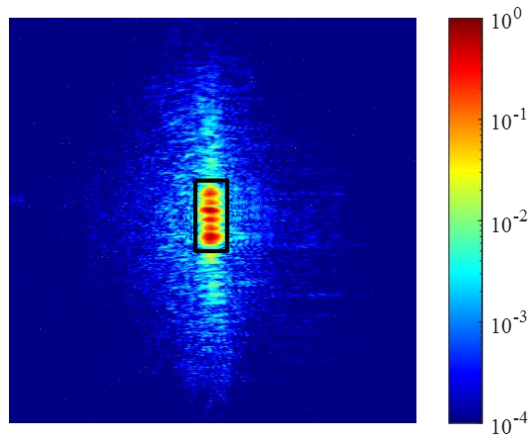

Fig. 4. Open loop output beam in log-scale. The black rectangle represents the 12 first Hermite-Gaussian modes spreading. Power outside of this rectangle is loss from imperfect MPLC component.

To summarize, the total beam combination efficiency including these three loss contributions, is about 70% if we subtract 0.5 dB loss from fiber connectors. This value is close to the one measured in [11] but with the advantage of having substantially more fibers, showing that the increase of the number of fibers does not dramatically affect the combination efficiency. Moreover, this efficiency his higher than the maximum achievable with tiled-aperture method but is lower than the highest value above 90% reported in [3] in a filled-aperture configuration. As way of improvement, insertion and scattering losses can be reduced with a higher quality device.

## 4. Higher order modes

Knowing the phases required to obtain the fundamental Gaussian mode and the transfer matrix of the MPLC, we can easily obtain the higher order modes output.

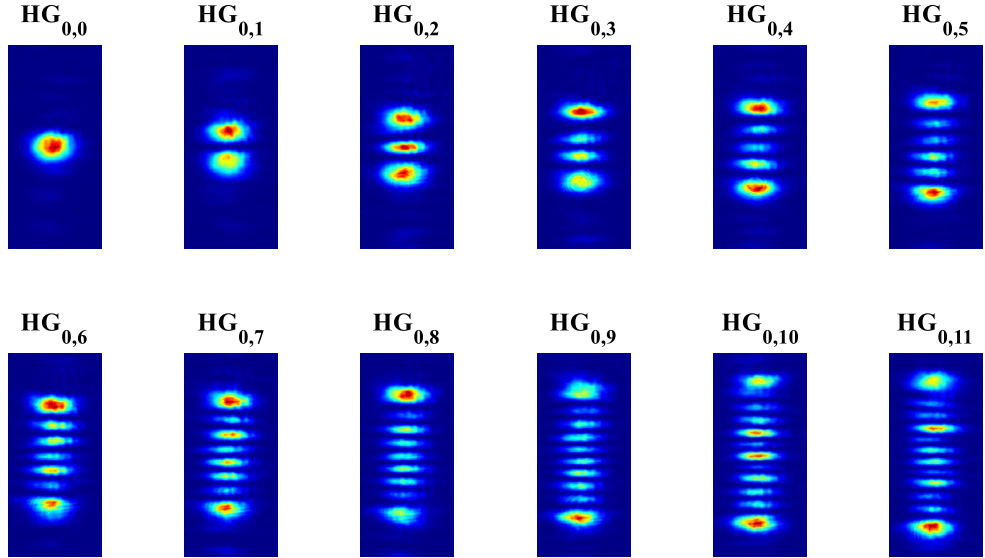

Fig. 5. Experimental Hermite-Gaussian output modes of our MPLC.

Fig. 5 represents the experimental output modes basis of our MPLC. The qualitative analysis based on the lobes number gives a good agreement with theory for all the modes. However, the differences with theory in lobe power distribution implies that the implemented transfer matrix is not exactly the theoretical one or that our Nelder-Mead algorithm did not converge exactly to the desired solution. The full mode characterisation, with phase and amplitude, is not straightforward and is out of the scope of this study.

## 5. Beam steering

A great advantage of using a MPLC for coherent beam combination is the ability to perform a continuous beam steering over the angular range of the highest order output mode. Fig. 6 describes the definition of the tilt of the fundamental HG mode. For sketch of simplicity, let us consider a one-dimensional transverse tilt from the optical axis, assuming that the other transverse component is Gaussian. The tilt corresponds to an angular displacement of the propagation axis by an angle $\theta$ and it is linked to the transverse momentum p of the beam by

$$p = \frac{2\pi \sin\theta}{\lambda} \approx \frac{2\pi\theta}{\lambda} \qquad (1)$$

in the limit of small angles and with λ the optical wavelength.

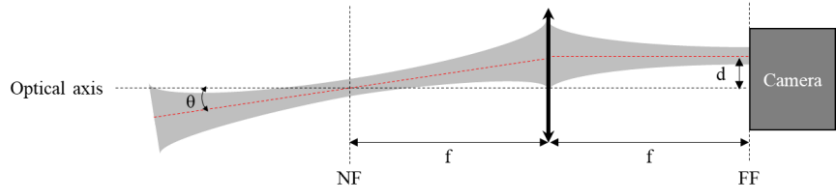

Fig. 6. Definition and measurement of the tilt. The tilt θ corresponds to an angular displacement of the propagation axis. The MPLC output waist defines the near field (NF) plane, and the tilt can be accessed in its far field (FF) by a displacement measurement d on the camera.

The transverse phase term then reads:

$$px = \frac{2\pi\theta}{\lambda}x = \theta_r x_r \qquad (2)$$

Where $x_r$ is the reduced coordinate $x_r = x\sqrt{2}/w_0$, $w_0$ being the waist of the fundamental mode, and $\theta_r$ is the reduced angle $\theta_r = \theta\sqrt{2}/\theta_0$, $\theta_0$ being the half diffraction angle of the fundamental Gaussian beam $\theta_0 = \lambda/\pi w_0$.

A tilted $E_\theta(x_r)$ Gaussian field is expressed as:

$$E_\theta(x_r) = HG_0(x_r)e^{i\theta_r x_r} \quad (3)$$

The Hermite-Gauss modes are eigenstates of the Fourier Transform, defined as:

$$f(x) \xrightarrow{FT} \tilde{f}(k) = \frac{1}{\sqrt{2\pi}} \int_{-\infty}^{+\infty} f(x)e^{-ikx} dx \quad (4)$$

Namely:

$$\widetilde{HG_n}(k) = (-i)^n HG_n(x) \quad (5)$$

The Fourier Transform of the tilted mode $E_\theta$ (near field) is then the displaced field $HG_0(k_r - \theta_r)$ (far field).

According to [13], a displacement $\theta_r$ of the fundamental mode can be decomposed on the higher order modes as follow:

$$HG_0(k_r - \theta_r) = \sum_{n=0}^{\infty} \alpha_n HG_n(k_r) \quad (6)$$

$$\alpha_n = \left(\frac{\theta_r}{\sqrt{2}}\right)^n \frac{1}{\sqrt{n!}} e^{-\frac{\theta_r^2}{4}} \quad (7)$$

Taking the inverse transform of expression eq. 6 and using eq. 5 and 7, the decomposition of a tilt on the HG basis mode gives:

$$E_p(x_r) = \sum_{n=0}^{\infty} \beta_n HG_n(x_r) \quad (8)$$

$$\beta_n = i^n \frac{\theta^n}{\theta_0^n \sqrt{n!}} e^{-\frac{\theta^2}{2\theta_0^2}} \quad (9)$$

If $T$ is the transfer matrix of the MPLC ($N$ individual fibers modes to $N$ HG modes), then the input fields to apply to the fibers are related to the aforementioned first $N$ $\beta_n$ coefficients by:

$$\vec{E}_{in} = T^t \vec{\beta} \quad (10)$$

The transfer matrix implemented in the MPLC we used is the Discrete Fourier Transform one. We see from eq. 10 that in order to keep a deflected beam with fundamental Gaussian mode, the system requires a full control of the input phases and amplitudes. However, for high power laser coherent beam combination application, the amplitude control is not desirable and only phase adjustment must perform the steering. That is why we have experimentally used the Nelder-Mead algorithm on the phase shifters to maximize the focused power at several steering angle.

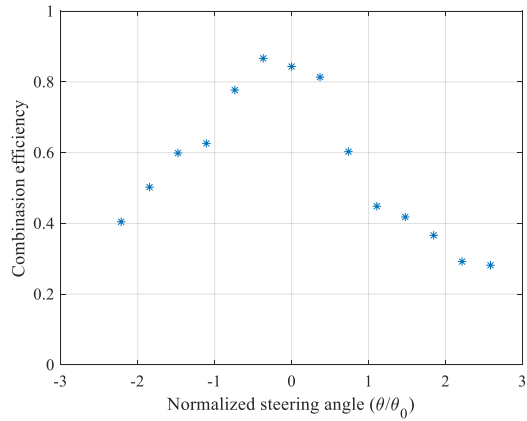

Fig. 7. Beam steering performance using the Nelder-Mead optimization. Combination efficiency from the power in the bucket at θ normalized by the theoretical power for a fundamental Gaussian beam.

Fig. 7 represents the results of the combination efficiency as a function of the normalized steering angle. The combination efficiency is defined as the ratio between the experimental powers in the bucket for one waist radius $w_0$ to the theoretical power for the Gaussian beam with lower divergence. This definition is well suited for a tilted beam that resembles a Gaussian beam since the efficiency reaches unity when the beam is exactly Gaussian. However it may overestimate the efficiency if the beam shape looks more like a top-hat adapted to the size of the photodiode. With this definition, the steering range where the combination efficiency is higher than 80%, is only one beam divergence and the efficiency decreases significantly after that.

Performances in beam steering with input phase control is highly sensitive to the choice of the transfer matrix implemented into the MPLC and the one implemented here results in low steering performances. An optimized matrix is under development and will lead to a high performances steering device with phase only control.

## 6. Conclusion

We performed the coherent beam combination of 12 fibers at 1.03 µm using a MPLC device and an interferometric phase lock loop. Even if the theoretical beam combination efficiency with this technology can be as high as 100%, we obtain experimentally only 70% combination efficiency. This low efficiency is partly due to Rayleigh scattering on phase plate and mirror material and can be overcome using higher performance dielectric coated optics. This device is also highly sensitive to alignment errors resulting in reduced combination efficiency. Using optimized components with high reflectivity and low scattering we expect up to 90% efficiency. We also obtained the higher order output Hermite-Gaussian modes with fair qualitative agreement to the theory. Finally, we tested the beam steering capability of the device and obtained limited performances due to a non-optimized transfer matrix of the MPLC device. Work on better matrices to perform high combination efficiency and to steer the beam with phase only control on the input fibers is in progress.